\documentclass[a4paper,11pt]{article}
\usepackage{pos}
\usepackage{multirow}
\DeclareUnicodeCharacter{03B3}{}

\title{Application of convolutional neural networks for data analysis
in TAIGA-HiSCORE experiment
}

\author[a]{Anna Vlaskina}
\author*[a,b]{Alexander Kryukov}

\affiliation[a]{M.V. Lomonosov Moscow State University, Leninskie Gory, Moscow, 119991, Russian Federation}

\affiliation[b]{Skobeltsyn Institute of Nuclear Physics, 1(2) Leninskie gory, Moscow 119991, Russian Federation}

\emailAdd{vlaskina.aa18@physics.msu.ru}
\emailAdd{kryukov@theory.sinp.msu.ru}

\abstract{The TAIGA experimental complex is a hybrid observatory for high-energy gamma-ray astronomy in the range from 10 TeV to several EeV. The complex consists of such installations as TAIGA-IACT, TAIGA-HiSCORE and a number of others. The TAIGA-HiSCORE facility is a set of wide-angle synchronized stations that detect Cherenkov radiation scattered over a large area. TAIGA-HiSCORE data provides an opportunity to reconstruct shower characteristics, such as shower energy, direction of arrival, and axis coordinates. The main idea of the work is to apply convolutional neural networks to analyze HiSCORE events, considering them as images. The distribution of registration times and amplitudes of events recorded by HiSCORE stations is used as input data. The paper presents the results of using convolutional neural networks to determine the characteristics of air showers. It is shown that even a simple model of convolutional neural network provides the accuracy of recovering EAS parameters comparable to the traditional method. Preliminary results of air shower parameters reconstruction obtained in a real experiment and their comparison with the results of traditional analysis are presented.}

\FullConference{%
  The 6th International Workshop on Deep Learning in Computational Physics \\
  July 6-8, 2022\\
  Dubna, Russia
}

\begin{document}
\maketitle

\section{Introduction}
In the high energy range (from $10^{15}$ eV), gamma rays and charged cosmic rays can be analyzed by studying the air showers they induce in the atmosphere. When the primary particle reaches the atmosphere and interacts with its atoms, it produces a cascade of secondary particles along the path of the primary cosmic ray emitting Cherenkov light. Such a cascade of particles is called \textbf{extensive air shower (EAS)}. The development of an extensive air shower depends on the characteristics of the primary particle: on its energy, the direction from which the cosmic ray came, and the type of particle. These parameters are of fundamental importance not only for studying extensive air showers, but also for studying processes using the methods of gamma astronomy and the physics of charged cosmic rays. \par
In this paper, we propose the use  of deep learning technologies to determine the energy of the primary particle and the direction of the EAS axis for events recorded by the TAIGA HiSCORE observatory - an array of Cherenkov detectors.

TAIGA (Tunka Advanced Instrument for cosmic rays and Gamma Astronomy) is a hybrid observatory for studying the physics of high and ultrahigh energy cosmic rays, located in the Tunka Valley, 50 km from Lake Baikal \cite{TAIGA}. This experiment is a complex system for ground-based astronomy from several TeV to several EeV. The observatory includes Cherenkov telescopes TAIGA-IACT, an array of wide-angle Cherenkov stations TAIGA-HiSCORE, an array of Cherenkov detectors Tunka-133, muon and electron detectors TAIGA-Muon and Tunka-Grande, which are an array of aboveground and underground scintillator counters, as well as detectors Tunka-REX radio emission. The use of several different types of detectors provides possibilities for multimessenger analysis of extensive atmospheric showers. \par
\textbf{TAIGA-HiSCORE} \cite{HiSCORE} is an array of wide-angle Cherenkov stations for studying cosmic rays and gamma-ray sources. The HiSCORE stations register EAS Cherenkov photons, as well as the signal arrival time. The advantages of the TAIGA-HiSCORE array is a large area for collecting statistics of about several kilometers and, at the same time, high sensitivity of the stations.

\section{ Deep learning method for TAIGA-HiSCORE experiment}
Reconstruction of the parameters of extensive air showers, the study of their composition and the energy of the primary particles are important problems in cosmic ray physics and gamma-ray astronomy. 
To reconstruct EAS parameters, we propose to use machine learning technologies, and, in particular, deep learning. Deep learning methods have shown their effectiveness in physical problems and, in particular, in problems of gamma astronomy. For example, in \cite{gammah}, convolutional recurrent neural networks were used to clean images from background noise, determine the EAS direction and the energy of the primary particle in Cherenkov telescope events. Deep learning methods were also used in the TAIGA experiment to analyze TAIGA-IACT events \cite{Postnikov}. Convolutional neural networks have shown their high efficiency in gamma-hadron separation and determining the energy of an EAS primary particle compared to traditional methods. However, unlike the arrays of pixels that make up the image of Cherenkov telescopes, the array of Cherenkov stations is sparse, since each of the stations is located a hundred meters from each other. Also, a distinctive feature of the problem under study is the ability to read both the amplitudes of the recorded signals and take into account the signal registration time. The novelty of the data processing of the TAIGA experiment in this work lies in the use of multimessenger analysis along with deep learning technologies.
In the proposed method, we present the arrays of data received from HiSCORE stations as images. The image may consist of signal registration time data or signal amplitude data. Working with the signal registration time data, we look for a time isoline and use it to reconstruct the direction of the shower axis. Deep learning technologies can be used to determine the time baseline. In particular, we propose the use of convolutional neural networks to reconstruct EAS parameters. 
\begin{figure}
\begin{minipage}[t]{0.5\linewidth}
\center{\includegraphics[scale=0.50]{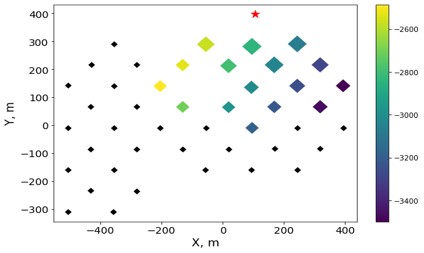} \\ a}
\end{minipage}
\hfill
\begin{minipage}[t]{0.5\linewidth}
\center{\includegraphics[scale=0.5]{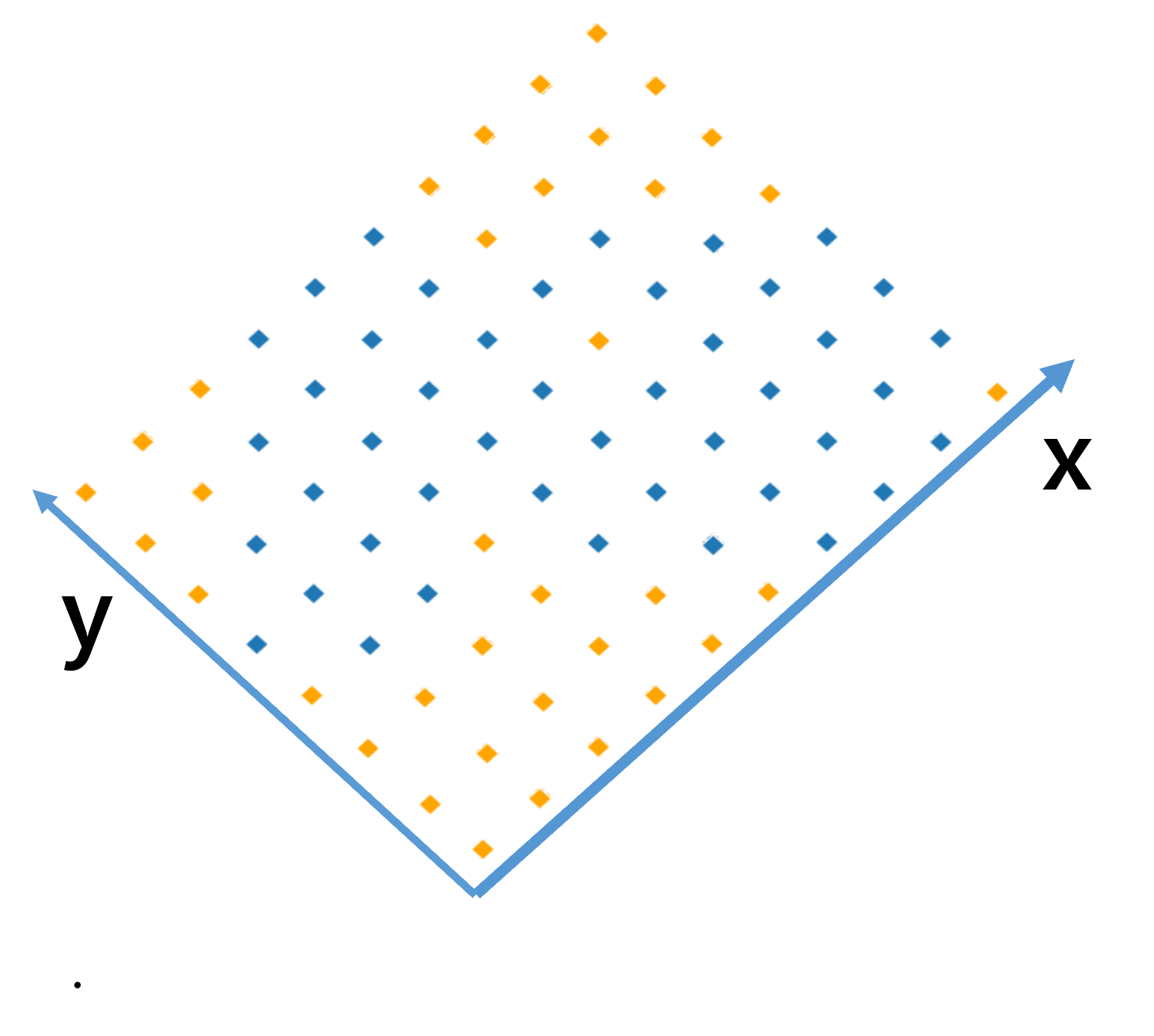} \\ b}
\end{minipage}
\caption{a: An example of an EAS event modeled for the HiSCORE array. The color indicates the time of registration of the particle, the value of the dot corresponds to the value of the signal amplitude. b: The array of station coordinates is reduced to a rectangular shape by adding points with zero values and rotating the coordinate axes by $45^{\circ}$. Added points are marked in orange. }
\label{fig:array}
\end{figure}
For training and validation during the learning process of the convolutional network, model data sets obtained using the CORSIKA program (COsmic Ray SImulations for KAscade) were used. This is software for detailed simulation of extensive air showers based on the Monte Carlo simulation \cite{CORSIKA}. To train and evaluate the performance of the model, the work uses Monte Carlo simulation data for an array of HiSCORE Cherenkov arrays from 2014. In addition to modeling EAS development, the responses of Cherenkov stations were simulated for the HiSCORE array for $44$ stations. The distance between "neighboring" \ detectors is approximately $106$ m. The data for each Cherenkov station that registered an event are information about the signal registration time, signal duration and its amplitude. Only those events in which at least four stations were triggered were selected. In total, the sample consists of 12216 events. 816 events were selected to evaluate the effectiveness of the model, the model was trained on 11400 events. \par
In order to take full advantage of the method of convolutional neural networks, which are specialized in local pattern recognition \cite{DeepL}, and achieve good stability and network optimization efficiency, the data is converted into formats suitable for the neural network. An EAS event registered by identical detectors can be considered as a pixelated image. In images, pixels usually form a Cartesian grid, so to bring the array to a rectangular shape, it suffices to rotate the coordinate axes of the stations by $45^{\circ}$. Since the input of the convolutional neural network is a two-dimensional array, the array of 44 stations is supplemented with points with zero values, as shown in Figure \ref{fig:array}. Thus, we get an array of dimensions $(10 \times 8)$. \par
The scheme of our CNN model is presented in Figure 2.
\begin{figure}[!ht]
\centering
\includegraphics[scale = 0.45]{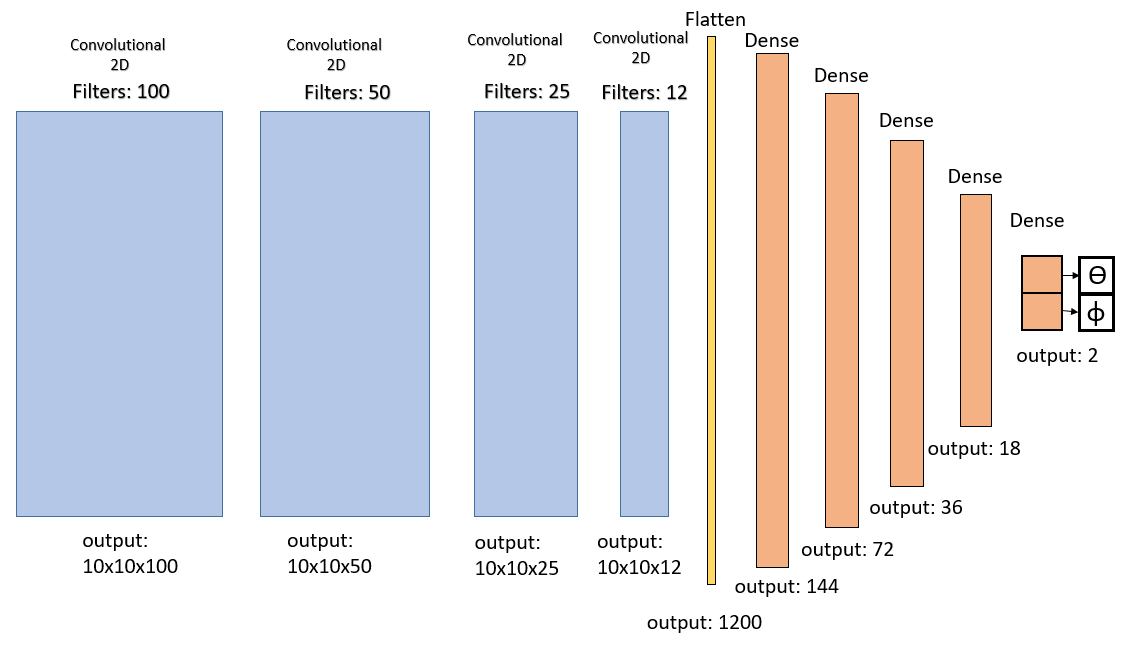}
\smallskip
\caption{The proposed neural network architecture. }
\end{figure}
Our model is a sequential convolutional neural network with four convolutional layers and six dense layers on top of it. The input layer contains 80 feature maps; on each subsequent convolutional layer, the number of feature maps is halved. In the convolutional layers feature extraction takes place, while in the dense layers linear regression is performed. 
Several model architectures have been tried, including those that use the pooling method and architectures with different kernel sizes, but the efficiency of such models was not higher than that proposed. We chose ReLU (Fig.\ref{fig:ReLU}.a) as an activation function on all layers since the values of angles are positive. Table 1 details the CNN parameters.  To implement the model, the work uses the TensorFlow \cite{TensorFlow} and Keras \cite{Keras}  libraries, used for machine learning on Python 3.1.1. This model could be adjusted to determine other parameters, such as energy and air shower axis coordinates.

\begin{figure}
\begin{minipage}[h]{0.45\linewidth}
\center{\includegraphics[scale=0.50]{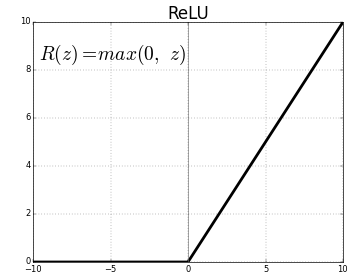} \\ a.}
\end{minipage}
\hfill
\begin{minipage}[h]{0.5\linewidth}
\center{\includegraphics[scale=0.35]{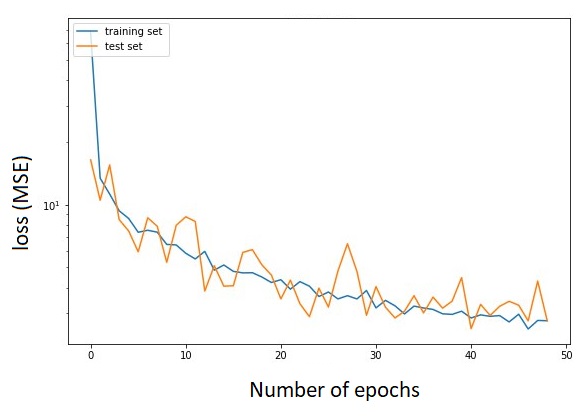} \\ b.}
\end{minipage}
\caption{a: Activation function ReLU b: Error function plot during model training for training and test sets }
\label{fig:ReLU}
\end{figure}

\begin{table}[h]
\centering
\caption{Model parameters.}
\medskip
\begin{tabular}{|c|c|c|}
\hline 
Kernel size & $2 \times 2$ \\
\hline
Learning rate & 0.001\\
\hline
Number of model parameters & 135,402 \\
\hline
Optimizer & ADAM (Adaptive Moment Estimation) \\
\hline 
Loss & $MSE = \frac{1}{n}\sum_i^n (y_i^{true} - y_i^{pred})^2 $ \\
\hline
Padding & 'same'\\
\hline 
Batch size & 15 events \\
\hline
Number of epochs & 50 \\
\hline
\end{tabular}
\end{table}

\section{Results}
The figure \ref{fig:ReLU}.b shows a plot of the error function during model training. Overall, the model demonstrated total loss of $MSE =  2.80 $ for two parameters. Let`s take a closer look at results for each parameter. We will estimate the deviations from the expected values using the mean absolute error function (MAE):
$$MAE = |y_{true} - y_{pred}|$$
The table \ref{tab:MAE} shows the average absolute errors of the zenith and azimuth angles.
\begin{table}[h]
\centering
\begin{tabular}{ |c|c|}
\hline
  EAS parameter & Mean absolute error \\
 \hline
 zenith angle $\theta$ & $0.97^{\circ}$ \\
 \hline
 azimuth angle $\phi$ & $1.38^{\circ}$ \\
 \hline

\end{tabular} 
\caption{Mean absolute errors of  zenith angle $\theta$ and azimuth angle $\phi$}
 \label{tab:MAE}
\end{table}
Errors for events with different numbers of triggered stations can vary significantly. Let's divide our dataset into events with the number of triggered stations more than ten and less than ten. Two parts of the sample contain 2684 and 9532 events, respectively. Mean absolute errors for events with number of triggered stations below ten and above ten are shown in the Table \ref{tab:below-above}: \par

\begin{table}[h]
\caption{Errors for events with the number of triggered stations less than 10 and more than 10:}
\label{tab:below-above}
\centering 
\begin{tabular}{|c|c|c|}

\hline
 & EAS parameter & Mean absolute error  \\
 \hline
  \multirow{2}{*}{Less than 10:} & zenith angle  $\theta$ & $1.34^{\circ}$ \\
 \cline{2-3}
 & azimuth angle $\phi$ & $1.68^{\circ}$ \\
 \hline
 \multirow{2}{*}{More than 10:}  & zenith angle  $\theta$ & $0.85^{\circ}$ \\
 \cline{2-3}
 & azimuth angle $\phi$ & $1.12^{\circ}$ \\
\hline
 
\end{tabular}\\

\end{table}

\begin{figure}[h]
\begin{minipage}[t]{0.5\linewidth}
\center{\includegraphics[scale=0.6]{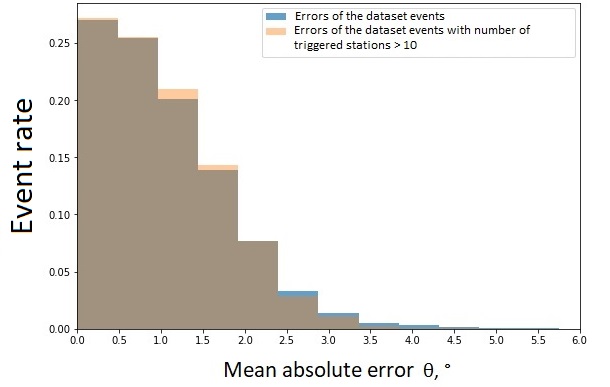} \\ a}
\end{minipage}
\hfill
\begin{minipage}[t]{0.5\linewidth}
\center{\includegraphics[scale=0.6]{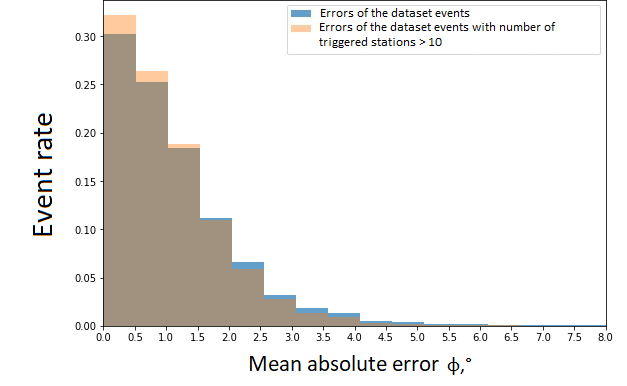} \\ b}
\end{minipage}
\caption{a: Error distribution for zenith angle $\theta$. b: Error distribution for azimuth angle $\phi$}
\label{fig:errors}
\end{figure}

The graphs on Figure \ref{fig:errors} show the distributions of errors in the azimuth and zenith angles in determining the EAS direction. 

\begin{figure}[h]
\begin{minipage}[t]{0.5\linewidth}
\center{\includegraphics[scale=0.62]{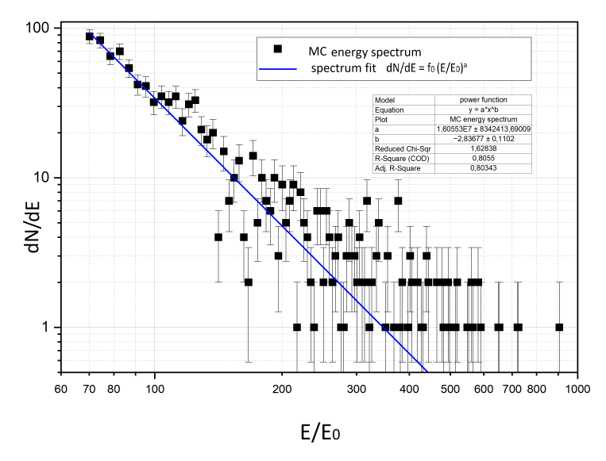} \\ a}
\end{minipage}
\hfill
\begin{minipage}[t]{0.5\linewidth}
\center{\includegraphics[scale=0.35]{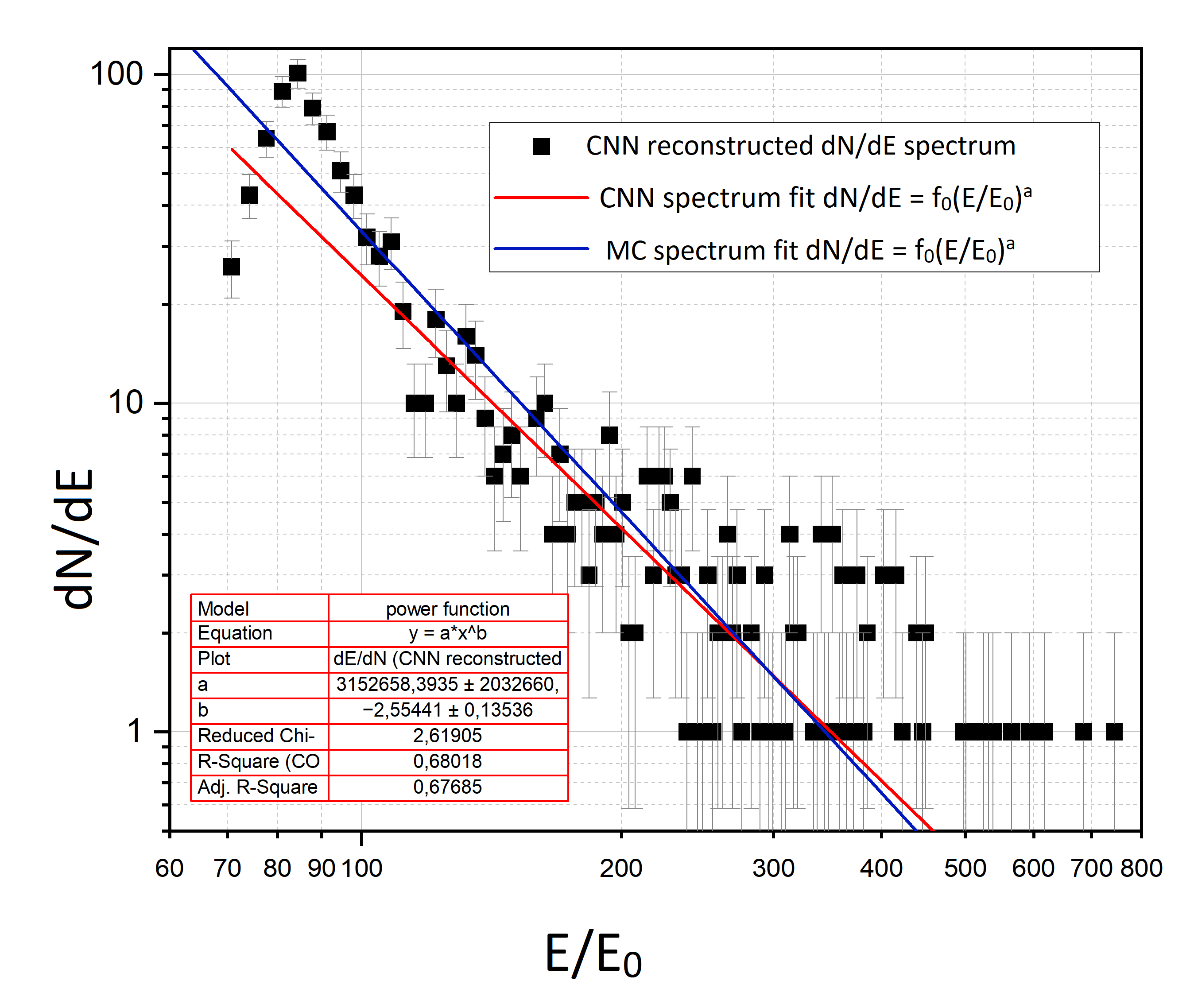} \\ b}
\end{minipage}
\caption{a: MC energy spectrum. b: CNN-reconstructed energy spectrum. The dotted red line corresponds to the approximation of points with energies greater than 90 TeV    }
\label{fig:distrib fit}
\end{figure}

The graphs on Figure \ref{fig:distrib fit} show the energy spectrum of Monte Carlo events that used to train the model and evaluate its performance, and the energy spectrum reconstructed by the neural network. The spectra were approximated by the following function: 
$$\frac{dN}{dE} = f_0 (\frac{E}{E_0})^a, E_0 = 1 \ TeV , $$ $f_0$-- normalization coefficient, the value of which depends on the experiment, while $a$ is a photon index, the value of which is most significant for the analysis of energy distribution. Distribution type taken from work \cite{Crab}

The Table \ref{tab:fit} shows the power-law parameters of the obtained spectra and the power-law parameter of the Crab Nebula spectrum fit from \cite{Crab}

\begin{table}[h]
\centering
\caption{}
\medskip
\begin{tabular}{|c|c|c|}
\hline 
$dN/dE $  spectra & a parameter & Red. Chi-Sq\\
\hline
MC spectrum & $-2.83 \pm 0.11$ & 1.628\\
\hline
CNN-reconstructed spectrum & $-2.55 \pm 0.14$ & 2.619 \\
\hline
MAGIC spectrum & $-2.47 \pm 0.01$ & 1.818\\
\hline

\end{tabular}
\label{tab:fit}
\end{table}

\section{Conclusions}
In this paper, we have demonstrated the possibility of using deep learning methods to determine the parameters of extensive air showers. The proposed model, which is a convolutional neural network, can be applied to determine the zenith and azimuth angles of the EAS axis. A deep learning method for energy spectrum reconstruction is also proposed. The trained convolutional neural network makes it possible to determine the zenith and azimuth angles of the EAS axis with an average accuracy of about 1 degree. The deep learning method shows good results in the reconstruction of the EAS energy spectrum - the slope of the spectrum reconstructed using a neural network is in good agreement with the MAGIC experiment.

\begin{acknowledgments}
The work was supported by the Russian Science
Foundation, grant №22-21-00442. The authors would like to thank the TAIGA collaboration for their support and the provision of the model data.
\end{acknowledgments}
\bibliographystyle{plain}
\bibliography{skeleton}

\end{document}